# A general approach to improve the bias stability of NMR gyroscope

Haifeng Dong, Min Hu

School of Instrumentation Science & Optoelectronics Engineering, Beihang University

**Abstract**

**In recent years, progress in improving the bias stability of NMR gyroscopes has been hindered. Taking inspiration from the core idea of rotation modulation in the strapdown inertial navigation system, we propose a general approach to enhancing the bias stability of NMR gyroscopes that does not require consideration of the actual physical sources. The method operates on the fact that the sign of the bias does not follow that of the sensing direction of the NMR gyroscope, which is much easier to modulate than with other types of gyroscopes. We conducted simulations to validate the method's feasibility.**

NMR gyroscope is used to detect the rotation of an object in the inertial reference frame. Detailed mechanism introduction of NMR gyroscope can be found in reference [1]. The basic measurement model is

$$w = \gamma B + \Omega \qquad (1)$$

where $w$ is the measured precession frequency, $\gamma$ is the gyromagnetic ratio, $B$ is the static field and $\Omega$ is the rotation angular rate of the object in the inertial reference frame.

It is obvious from equation (1) that the variation of static field will cause the bias unstability of the gyroscope output. To avoid this problem, two nuclear species are often used in one vapor cell. One is used to measure the bias field in-situ and compensate it in real time. The other is used to output the rotation angular rate insensitive to the variation of static field[2]. In this way, we have two equations,

$$\begin{aligned} w_1 &= \gamma_1 B + \Omega \\ w_2 &= \gamma_2 B + \Omega \end{aligned} \qquad (2)$$

Where the subscript 1 and 2 represent nuclear one and nuclear two, for example, 129Xe and 131Xe, respectively.

If $w_1$ is stabilized, then

$$w_2 = \left(1 - \frac{\gamma_2}{\gamma_1}\right)\Omega + \frac{\gamma_2}{\gamma_1} C \qquad (3)$$

Where $C = w_1$ is a constant due to the feedback mechanism.

Equation (3) is completely equivalent to the traditional two nuclear compensation model, i.e. the solution of equation (2),

$$\Omega = \frac{\gamma_1 w_2 - \gamma_2 w_1}{\gamma_1 - \gamma_2} \qquad (4)$$

However, equation (3) provides a basis for deciding which nuclear should be used to compensate the static field according to the scale factor enhancement, i.e. $1-\gamma_2/\gamma_1$.

The real case is more complex than equation (2) as the alkali has to be used to spin-exchange pump the nuclear and to probe the nuclear precession. Due to Fermi-contact hyperfine interaction, nuclear Larmor frequency is shifted by the alkali spin polarization[1]. As the shift ratio between two nuclear is not equal to $\gamma_2/\gamma_1$, it is an anomaly field similar to object rotation. In this way, the alkali polarization become an important bias instability source because many factors can affect the polarization, such as temperature, laser frequency and amplitude, and the relaxation.

There are some ways to avoid the bias instability caused by variation of alkali polarization. One is synchronous spin-exchange pumping, which make the alkali polarization vertical to the equivalent static field, thus the variation of total field caused by the alkali polarization becomes a second-order small quantity[3-5]. The second is depolarizing alkali spins by rf field during free precession intervals and minimizing Rb polarization along the bias field during the probe intervals[6, 7]. The third is utilizing the PT transition to extends the dual species NMR gyroscope to a three-component comagnetometer[8]. All of these methods provide important contribution to the bias stability of NMR gyroscopes.

Here, inspired by rotating modulation in the strapdown inertial navigation system[9], we proposed a general method that can improve the bias stability for all the types of NMR gyroscopes. In the traditional rotation modulation, the gyroscope is rotated mechanically. Fortunately, for NMR gyroscope, this can be easily realized by changing the direction of the bias field, no matter which is a static field[6-8] or average of pulse field[3-5].

Although there are no static field $B$ in Equation (3) and (4), we notice that the direction of the measurement is still decided by $B$, because once $B$ is negative, the equation (2) should be written as,

$$w_1 = \gamma_1 B - \Omega$$
$$w_2 = \gamma_2 B - \Omega \qquad (5)$$

And Equation (3) and (4) become,

$$w_2 = -\left(1-\frac{\gamma_2}{\gamma_1}\right)\Omega + \frac{\gamma_2}{\gamma_1}C \qquad (6)$$

and

$$\Omega = -\frac{\gamma_1 w_2 - \gamma_2 w_1}{\gamma_1 - \gamma_2} \qquad (7)$$

, respectively. Equations (6) and (7) each feature a negative sign that is absent from equations (3) and (4). From the point view of phase detection, the output of phase $\phi$ is proportional to the $\Delta w$, the sign of which also changes with that of the static field[10].

The static field in an NMR gyroscope is typically generated using coils, and its direction can be readily altered by reversing the current flow. A viable approach to achieve this is by utilizing a single pole double throw switch, such as the SA630, that is manufactured using CMOS technology, or by employing a MEMS process that incorporates a mechanical contact[11].

The output of gyroscope after square waveform modulation become,

$$S_{t1} = k \cdot \Omega + Bias$$
$$S_{t2} = -k \cdot \Omega + Bias$$
$$S_{t3} = k \cdot \Omega + Bias \quad (8)$$
$$S_{t4} = -k \cdot \Omega + Bias$$
$$\ldots\ldots$$

where $S$ means the measurement signal, t1, t2, t3, t4… present the time interval of the measurement, $k$ is the scale factor, $\Omega$ is the object rotation angular rate and $Bias$ is the bias of the gyroscope which drifts with time.

By integration, demodulating or just differentiate the adjacent recording, most of the long-term bias drift can be eliminated. The essential idea is just the same with rotating modulation. Considering that the sensing direction of NMR gyroscope can be controlled electrically, this method has great advantage in enhancing the bias stability of NMR gyroscope without the need for additional cost and/or facilities. Furthermore, the modulation frequency can be higher and the modulation method can be more diverse than mechanical modulation.

We conducted simulations, and the outcomes are presented in Fig. 1 and Fig. 2, respectively. The amplitude of phase of 131Xe is amplified by a factor of 20, which is related to the pumping rate. The x-axis represents time in seconds, while the y-axis represents the signal amplitude in arbitrary units. At 100s, the feedback switch was closed, and at 250s, the field was reversed. Fig. 1 depicts the simulation results when the rotation input was zero, and Fig. 2 represents the case when the rotation input was 0.001rad/s. The offset in Fig. 1 was due to the alkali polarization variance, which had the same direction for the reversed field. As previously analyzed, the field reversal led to a corresponding reversal of the rotation output.

We observe an interesting phenomenon that the drift due to variation of pumping rate Rp for reverse field is different, i.e. there is a factor. If this factor can be accurately calibrated, then the drift can be eliminated more efficiently.

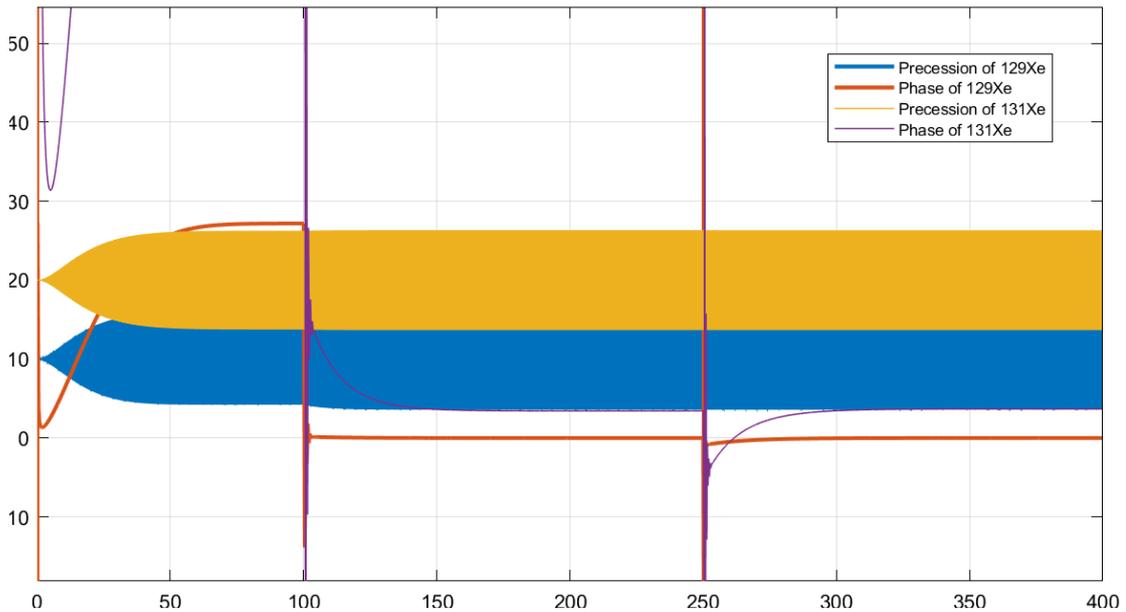

Fig. 1 The output of NMR gyroscope when the rotation input is zero

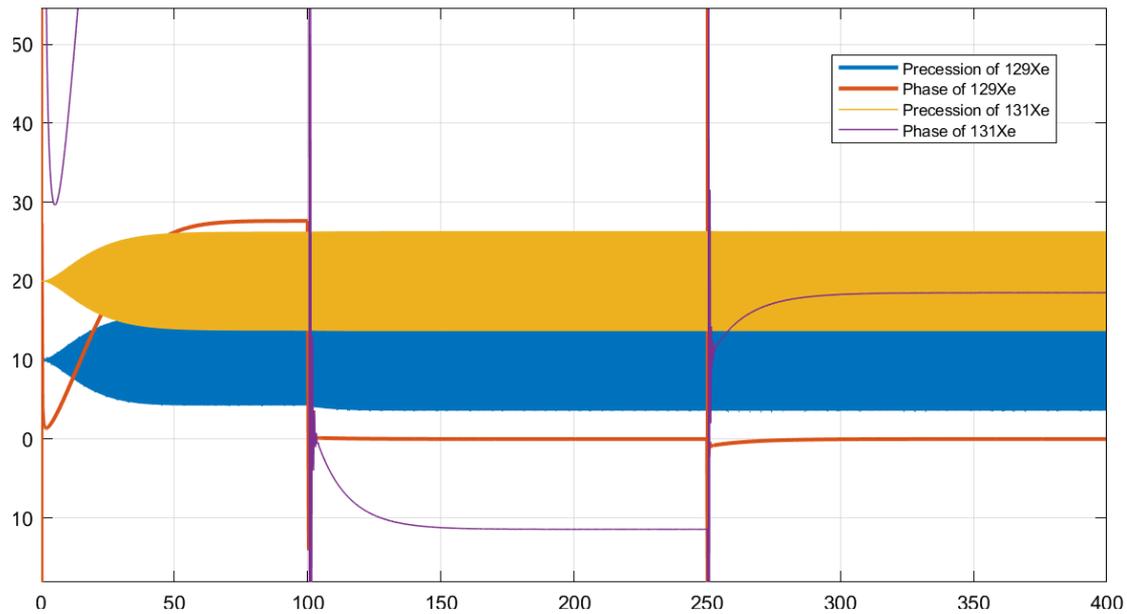

Fig. 1 The output of NMR gyroscope when the rotation input is 0.001rad/s

Based on the NMR sensing mechanism and inspired by the key concept of the rotation modulation in strapdown navigation system, we have proposed a way to enhance the bias stability of NMR gyroscope. Besides NMR gyroscope, the method can actually be used for any other gyroscopes that can alternate the sensing direction easily. Furthermore, as the rotation modulation has been developed for many years[9], there may be many improvement ways for NMR gyroscope draw on related methods.

**Reference:**

[1]　T. G. Walker and M. S. Larsen, "Spin-exchange-pumped NMR gyros," in *Advances in atomic, molecular, and optical physics*, vol. 65: Elsevier, 2016, pp. 373-401.

[2]　H. Dong and Y. Gao, "Comparison of Compensation Mechanism Between an NMR Gyroscope and an SERF Gyroscope",　*IEEE Sensors Journal,* Vol. 17, No. 13, pp. 4052-4055, 2017.

[3]　A. Korver, D. Thrasher, M. Bulatowicz and T. G. Walker, "Synchronous Spin-Exchange Optical Pumping",　*Phys Rev Lett,* Vol. 115, No. 25, p. 253001, Dec 18 2015.

[4]　D. Thrasher *et al.*, "Continuous Comagnetometry using Transversely Polarized Xe Isotopes", *Pysical Review A,* Vol. 100, p. 061403, 2019.

[5]　Thad G. Walker, S. S. Sorensen and Z. S. Yardim, "Dual Isotope NMR Gyro",　*Procding of SPIE,* 2021.

[6]　D. Sheng, A. Kabcenell and M. V. Romalis, "New Classes of Systematic Effects in Gas Spin Comagnetometers",　*Physical Review Letters,* Vol. 113, No. 16, p. 163002, 2014.

[7]　M. E. Limes, D. Sheng and M. V. Romalis, "3He-129Xe Comagnetometery using 87Rb Detection and Decoupling",　*Phys Rev Lett,* Vol. 120, No. 3, p. 033401, Jan 19 2018.

[8]　X. Zhang, J. Hu and N. Zhao, "Stable Atomic Magnetometer in Parity-Time Symmetry Broken Phase",　*Phys Rev Lett,* Vol. 130, No. 2, p. 023201, Jan 13 2023.

[9]　J. Feng, "A Review of Rotary Modulation Technology Applied to Strapdown Inertial Navigation System," in *2018 IEEE CSAA Guidance, Navigation and Control Conference (GNCC)*, 2018.

[10]　D. Budker and D. F. J. Kimball, *Optical Magnetometry*. Cambridge University Press, 2013.

[11]　M. Tang, A. Q. Liu and A. Agarwal, "A Low-Loss Single-Pole-Double-Throw (SPDT) Switch

Circuit," in *Solid-state Sensors, Actuators & Microsystems Conference, Transducers International*, 2007.